\documentclass[journal,twoside,web]{ieeecolor}
\usepackage{tmi}
\usepackage{cite}
\usepackage{soul}
\usepackage{amsmath,amssymb,amsfonts}
\usepackage{algorithmic}
\usepackage{graphicx}
\usepackage{textcomp}
\usepackage{tabularx}
\usepackage[super]{nth}
\usepackage{breqn}
\usepackage{float}
\usepackage[super]{nth}
\usepackage[separate-uncertainty=true,parse-numbers=false]{siunitx}
\usepackage{todonotes}
\def\BibTeX{{\rm B\kern-.05em{\sc i\kern-.025em b}\kern-.08em
    T\kern-.1667em\lower.7ex\hbox{E}\kern-.125emX}}
\usepackage{multirow}
\usepackage{tikz}
\usepackage{comment}
\def\checkmark{\tikz\fill[scale=0.4](0,.35) -- (.25,0) -- (1,.7) -- (.25,.15) -- cycle;} 
\usepackage{xcolor}
\sethlcolor{yellow}
\definecolor{Myred}{HTML}{C00000}
\newcommand{\rev}[1]{{\color{black}#1}} 
\newcommand{\rerev}[1]{{\color{black}#1}} 
\usepackage[switch,columnwise]{lineno}

\graphicspath{{../figures/}}

\begin{document}
	

\title{Quantitative ultrasound imaging of bone: anatomical images, tissue structural quality, and pulsatile blood flow }

\author{Gabrielle Laloy-Borgna, Nastassia Navasiolava, Pim Hutting, Andréa Bertona, Amadou S. Dia, Sébastien Salles, Anthony Augé, Alice Mazzolini, Quentin Grimal, Olivier Lucidarme, Hervé Locrelle, Jacques-Olivier Fortrat, Laurence Vico, Marc-Antoine Custaud and Guillaume Renaud.
\thanks{This work was supported in part by SATT LUTECH under grants BONES and WELLBONES. \textit{(Corresponding authors: Gabrielle Laloy-Borgna and Guillaume Renaud)}}
\thanks{Gabrielle Laloy-Borgna, Guillaume Renaud and Pim Hutting are with the
Department of Imaging Physics, Delft University of Technology, Delft, The Netherlands (e-mail: g.laloy-borgna@tudelft.nl, g.g.j.renaud.@.tudelft.nl and p.r.p.hutting@student.tudelft.nl).}
\thanks{Nastassia Navasiolava, Andréa Bertona, and Marc-Antoine Custaud are with the University of Angers, CRC, CHU Angers, Inserm, CNRS, MITOVASC, Équipe CARME, SFR ICAT, F-49000, Angers, France (e-mail: nastassia.navasiolava@chu-angers.fr, bertona.andrea@gmail.com, and macustaud@chu-angers.fr).}
\thanks{Jacques-Olivier Fortrat is with Vascular medicine unit, CHU Angers,  Inserm, CNRS, MITOVASC, Equipe CARME, SFR ICAT, F-49000, Angers, France (e-mail: jacques-olivier.fortrat@chu-angers.fr).}
\thanks{Laurence Vico and Hervé Locrelle are with INSERM, University Jean Monnet, Mines   Saint-Etienne, U1059 Saint Etienne, France (email: vico@univ-st-etienne.fr and herve.locrelle@chu-st-etienne.fr).}
\thanks{Amadou S. Dia, Sébastien Salles, Anthony Augé, Alice Mazzolini, Quentin Grimal and Olivier Lucidarme are with Laboratoire d'Imagerie Biomédicale, Sorbonne Université, Paris, France (email: sebastien.salles@cnrs.fr, anthony.auge@centrale.centralelille.fr, alice.mazzolini.am@gmail.com, quentin.grimal@sorbonne-universite.fr, olivier.lucidarme@aphp.fr).}
\thanks{Sébastien Salles is also with University of Bordeaux, Centre National de Recherche Scientifique (CNRS), Centre de Résonance Magnétique des Systèmes Biologiques (CRMSB), Bordeaux, France (email: sebastien.salles@cnrs.fr).}}

\maketitle

\begin{abstract}
We propose an ultrasound approach which provides, with one single examination and one single device, access to three bone biomarkers: anatomy, tissue quality and blood flow. It unlocks ultrasound imaging inside bone by accounting for ultrasound wave speed heterogeneity and anisotropic wave refraction. This study reports the first \emph{in vivo} evaluation with a comparison to peripheral Quantitative Computed Tomography (pQCT) and modulations of blood flow. 
Anatomical multi-layer bone-corrected reconstruction was validated at the tibia of healthy volunteers against pQCT and showed agreement on bone cortex interfaces. Estimation of axial and radial ultrasound wave speeds in cortical bone tissue (i.e. along the tissue symmetry axis and normal to it) demonstrated good reproducibility and positive correlation with bone mineral density measured by pQCT. Pulsatile blood flow was mapped and quantified in cortical and medullary regions. A directional ray selection method was developed to enhance blood signal extraction by reducing strong specular reflections originating from the outer and inner surfaces of the bone cortex. Physiological and non-physiological modulations of blood flow, namely head-up/head-down tilt table maneuvers and arterial occlusions, demonstrated the method sensitivity to blood flow variations. For the first time, reactive hyperemia was observed inside bone cortex. These results demonstrate the feasibility of a portable, non-ionizing, and quantitative ultrasound approach for structural, anatomical, and vascular characterization of bone tissue. This approach may offer new diagnostic capabilities for bone disorders, for instance osteoporosis, delayed fracture healing or osteonecrosis. 

\end{abstract}

 
\section{Introduction}
\label{sec:introduction}
\IEEEPARstart{A}{s} with other organs, bone health can be assessed by detecting and monitoring abnormalities in anatomy, tissue structure and blood flow.
Bone anatomy is currently most often assessed with x-ray imaging (plain radiography and computed tomography). Bone tissue quality is most often evaluated with Dual-energy X-ray Absorptiometry (DXA). DXA measures the areal bone mineral density, and is used for the diagnosis of osteoporosis \cite{yang_distribution_2014}. Unlike other organs, pulsatile blood flow in bone is currently out of reach. However, Magnetic Resonance Imaging and Positron Emission Tomography can detect regions with elevated (for instance bone marrow lesion) or reduced (osteonecrosis) time-averaged blood flow \cite{dyke_noninvasive_2010}. Cortical thickness and Bone Mineral Density (BMD) are linked to the fracture risk~\cite{chevalley_fracture_2013,yang_distribution_2014}, while intraosseous blood flow \rev{is, like for all other organs, linked to the metabolic activity. More specifically, in normal functioning, blood flow plays a central role in bone growth, remodeling and reparation. In addition, some bone disorders are due to hypo- or hyper-vascularization: osteonecrosis and delayed fracture healing are caused by ischemia in bone, while bone marrow lesions correspond to regions of higher blood flow~\cite{tomlinson_skeletal_2013, ramasamy_blood_2016, cowin_blood_2015}.} Yet, there is no non-invasive method able to map and characterize intra-osseous blood flow in the bone cortex.

Ultrasound imaging offers an attractive solution thanks to its portability, real-time capability, and absence of ionizing radiation. Although bone was long considered unsuitable for ultrasound due to its high acoustic impedance and attenuation, recent advances in beamforming have changed this perspective. However, by taking into account ultrasound wave speed heterogeneities, recent advances in beamforming and speed-of-sound correction make it possible. By accounting for spatial heterogeneities in ultrasound wave speed, ultrasound now enables the reconstruction of anatomical images not only of the bone tissue itself~\cite{renaud_vivo_2018}, but also of structures located behind strongly aberrating layers such as the skull~\cite{mozaffarzadeh_refraction-corrected_2022,bureau_three-dimensional_2023}. In addition, ultrasound wave speed inside bone has been shown to be a surrogate for tissue elasticity, density~\cite{granke_change_2011} and mechanical strength \cite{peralta_bulk_2021}. Furthermore, Doppler ultrasound, when combined with dedicated reconstruction and blood extraction methods, can detect blood flow within and behind the cortical bone~\cite{salles_revealing_2021}.

In this study, we demonstrate how a single ultrasound examination can yield three biomarkers at the diaphysis of long bones: (i) anatomical images corrected for refraction and anisotropy ; (ii) characterization of cortical bone tissue through the estimation of ultrasound wave speed and its anisotropy ; and (iii) imaging and quantification of pulsatile blood flow within both the bone cortex and the medullary cavity. Anatomical images and wave speed measurements are validated \emph{in vivo} against peripheral Quantitative Computed Tomography (pQCT), while blood flow imaging is evaluated through its sensitivity to modulations. To enhance blood signal extraction, a directional ray selection method was developed to suppress strong specular reflections originating from bone cortex interfaces. The sensitivity of this approach is demonstrated through two protocols: arterial occlusion and head-up/head-down tilt experiments.

\section{Bone-corrected image reconstruction}
All three biomarkers—anatomy, bone tissue quality, and blood flow—are derived from the same ultrasound acquisitions, processed through a unified image reconstruction framework. In this section, we describe how anatomical images are reconstructed by accounting for the effect of cortical bone on ultrasound propagation, including wave refraction and wave speed anisotropy.

\subsection{Methods for sequential multi-layer reconstruction}
\label{subsec:reconstruction}
\subsubsection{Ultrasound sequences}
To obtain high-quality anatomical images and accurately detect bone interfaces, a synthetic aperture (SA) sequence was first used. In this sequence, each element transmitted individually while all elements received, resulting in one transmission per element. \rev{This sequence achieves the best possible image quality, provided the Signal-to-Noise Ratio (SNR) is sufficient. However,} for blood flow detection, \rev{since the signal back-scattered by the red blood cells is weak and} a high frame rate is required, a second sequence was used based on the transmission of unfocused waves \rev{(plane or diverging waves)} with full aperture, \rev{yielding a better SNR.}
 
\subsubsection{Phase aberration correction}
In this work, the imaged medium was modeled as a series of homogeneous layers with distinct ultrasound wave speeds. Refraction at the interfaces was accounted for using ray-tracing methods. 
The silicone rubber lens at the front surface of the probe was modeled as the first layer in the medium. 

A sequential approach was used to construct an image inside bone. At first, the medium was modeled as homogeneous cutaneous tissue (wave speed: \qty{1580}{\metre\per\second}), and a two-layer ray tracing (lens \& skin) was performed. Using Djikstra's \cite{dijkstra_note_1959} algorithm, the outer bone surface (periosteum) was segmented. Once the periosteum was identified, a third layer corresponding to the cortical bone was added. The image was then reconstructed using a three-layer ray tracing model (lens, skin and cortical bone), which allowed segmentation of the inner cortical interface (endosteum). A fourth layer representing the marrow (medullary cavity) could be added.

Cortical bone was modeled as a medium with weak transverse isotropic elasticity, assuming axial symmetry of the wave speed. In this model, the wave speed depends on three parameters: the radial wave speed $C^{radial}$, the axial wave speed $C^{axial}$, and a non-dimensional anisotropy shape   coefficient $\alpha^{aniso}$ \cite{renaud_vivo_2018  }. Radial and axial wave speeds are determined by two elastic coefficients ($C_{11}$ and $C_{33}$) and the mass density ($\rho$): $C^{radial} = \sqrt{\frac{C_{11}}{\rho}}$ and $C^{axial} = \sqrt{\frac{C_{33}}{\rho}}$ \cite{granke_change_2011}. The wave speed in cortical bone is then given by: 
\begin{dmath}
	C^{bone}(\theta) = C^{axial} - (C^{axial}-C^{radial})\times (\alpha^{aniso} \sin ^2 \theta \cos ^2 \theta + \sin ^4 \theta)
\end{dmath}
where $\theta$ is the angle between the ultrasound ray and the symmetry axis of bone tissue micro-structure, which was assumed here parallel to the periosteum, based on histological evidence \cite{mccarthy_physiology_2006}.

\subsection{Validation of the approach on simulated data}
\begin{figure}[htb]
	\centering
	\includegraphics[width=\linewidth]{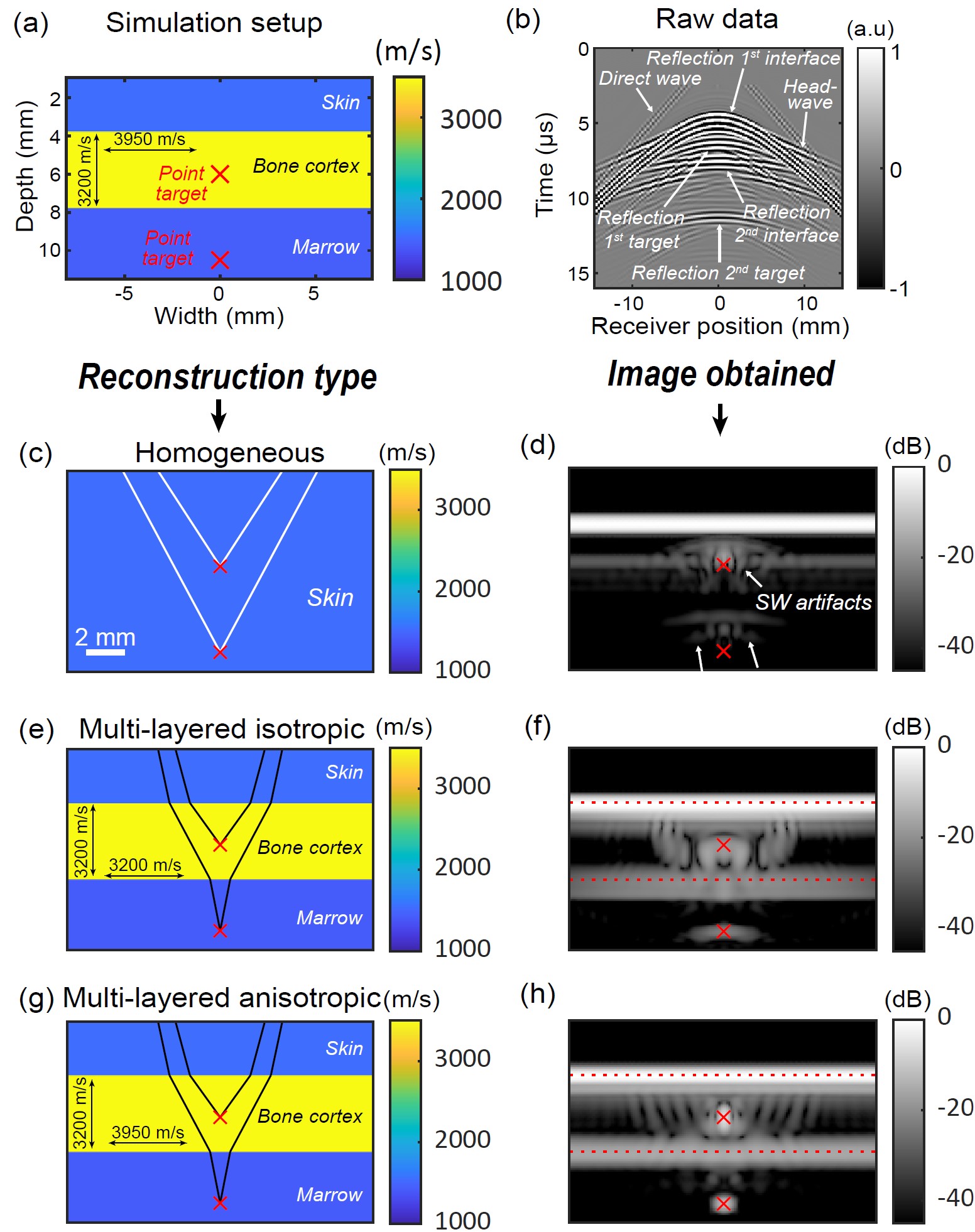}
	\caption{Simulated point spread functions (PSF) in a medium mimicking a longitudinal view of the tibia diaphysis. (a) Simulation setup showing the simulated medium containing two point targets. \rev{(b) Raw data for transmission with element \#48, plotted as a function of time and the receivers positions.} (c) Homogeneous ray tracing scheme and (d) corresponding B-mode image obtained. \rev{Artefacts resulting from mode-converted shear waves (SW) are pointed with white arrows.} (e) Multi-layered ray tracing assuming an isotropic bone layer and (f) corresponding image. (g) Multi-layered ray tracing assuming an anisotropic bone layer and (h) corresponding image.}
	\label{Fig-Methods}
\end{figure}

To evaluate the phase-aberration correction pipeline, a dataset was simulated using \textit{SimSonic 2D} \cite{bossy_simsonic_nodate,bossy_three-dimensional_2004}. \rev{2D simulations were considered sufficient given the geometry of the measurement site and the array dimensions of the ultrasound probes (22-28 mm wide and approximately 6-12 mm high). Indeed, the mid-diaphyseal portion of the tibia is approximately invariant by translation along its longitudinal axis ; therefore, in a transverse view, out-of-plane wave propagation is negligible. In addition, the probe was positioned onto the medial side of the diaphysis, which is locally nearly flat, so that in a longitudinal view, all bone reflections are contained within a plane. Finally, real-time ($\sim$10 images/second) image reconstruction with aberration correction provided continuous feedback to the operator, ensuring optimal probe positioning all along the acquisition.} The simulated setup mimicked the experimental configuration \rev{of the anatomical images of study 3}. It consisted \rev{of a P4-1 probe placed on top of a medium made }of a skin layer (\qty{1550}{\metre\per\second}) over a bone layer with anisotropic wave speed (between 3200 and \qty{3950}{\metre\per\second}, with $\alpha^{aniso}=1.4$) and a marrow layer (\qty{1430}{\metre\per\second}). The values used for these parameters were based on previous studies involving young healthy males~\cite{renaud_vivo_2018,renaud_measuring_2020}. Two point targets were inserted, one in the bone layer and the other one in the marrow layer (see Fig.~\ref{Fig-Methods}(a)). The imaging sequence \rev{was made of single-element transmissions, meaning each of the 96 probe elements was transmitting one-by-one, while all elements were receiving back-scattered signals. Raw data is shown in Fig.~\ref{Fig-Methods}(b) for element \#48 shooting, showing reflections from the two interfaces and the two point targets.}

The dataset was first reconstructed assuming a homogeneous medium using the full aperture in receive (Fig.~\ref{Fig-Methods}(c)). This produced low-contrast and distorted Point Spread Functions (PSF), shifted 1.9 and \qty{2.3}{\milli\meter} above their true locations (Fig. \ref{Fig-Methods}(d) and Table~\ref{Table-PSF}). \rev{In addition, shear wave artifacts resulting from mode conversion in the bone cortex layer could be identified underneath the PSF (see white arrows in Fig.\ref{Fig-Methods}(d))}. Then, the ray-tracing algorithm was applied using a three-layer model (skin, bone and marrow). The bone layer was initially considered isotropic (Fig. \ref{Fig-Methods}(e-f)). The resulting PSF showed improved localization, contrast, and full-width at half maximum (FWHM). Finally, wave speed anisotropy in cortical bone was considered (Fig. \ref{Fig-Methods}(g-h)). The PSF were correctly positioned while contrast and FWHM were further enhanced compared to the isotropic case. The axial shift, relative amplitude and FWHM for each method are reported in table~\ref{Table-PSF}. For the anisotropic case, the FWHM is in agreement with the theoretical resolutions (\num{0.60} and \qty{0.93}{\milli\metre}). \rev{It can also be noticed that since the PSF is well focalized in the last case (Fig.\ref{Fig-Methods}(h)), shear wave artifacts are barely visible because their amplitude is now negligible compared to the main lobe.}

These results demonstrate that the ray-tracing algorithm can efficiently correct phase aberrations caused by a cortical bone layer when wave speed anisotropy is taken into account.

{
	\setlength\arrayrulewidth{0.3pt}
\begin{table}[ht]
 	\caption{PSF features obtained from the simulation study for different reconstruction assumptions}
 	\label{Table-PSF}
 	\setlength{\tabcolsep}{2pt}
 	\renewcommand{\arraystretch}{1.2} 
 	\begin{tabularx}{\linewidth}{ |>{\centering\arraybackslash}p{4cm} 
 			| >{\centering\arraybackslash}p{1.7cm} 
 			| >{\centering\arraybackslash}p{1.2cm}  |  >{\centering\arraybackslash}p{1.3cm} |}
 			\hline
 			\multirow{2}{*}{\textsc{Reconstruction method}} & \textsc{Axial error} & \textsc{Max PSF}  & \textsc{FWHM} \\
 			
 			 & (mm) & (dB) & (mm)\\
 			\hline
 			\multirow{2}{*}{\textbf{Homogeneous reconstruction}}& -1.9 & -26 & 3.2 \\  
 			&-2.3 &-20 & 4.8 \\		\hline
 			\multirow{2}{*}{\textbf{Isotropic 2-layer ray tracing}} & 0.09 & -11 & 2.6 \\	&0.05 &-13 & 3.2 \\			\hline
 			\multirow{2}{*}{\textbf{Anisotropic 2-layer ray tracing}} & 0.09 & 0 & 0.63
 			 \\
			&0.05 &0 & 0.87 \\			\hline 			 
 			
 		\end{tabularx}
 	\end{table}
}

\subsection{\emph{In vivo} evaluation of the proposed framework: anatomical imaging of the tibia}

\subsubsection{Methods}

\paragraph{Study design}
This study aimed to acquire ultrasound data at the tibial diaphysis of healthy volunteers (Study 2, see table~\ref{Table-StudiesDesign}) and to compare ultrasound images to peripheral Quantitative Computed Tomography (pQCT) scans. To ensure variability in terms of bone quality and anatomy, 16 healthy male volunteers were recruited, divided into two age groups of 8 persons: 20-40 and 50-70 years old. All volunteers underwent both pQCT and ultrasound acquisitions at the tibial diaphysis. \rev{The study was promoted by CHU Angers, the approval number by appropriate ethical committee was 23.03951.000199-MS01 and the clinical trial number was NCT06206031.}


\begin{table}[ht]
	\caption{Studies on healthy volunteers used in this paper}
	\label{Table-StudiesDesign}
	\setlength{\tabcolsep}{3pt}
	\renewcommand{\arraystretch}{1.2} 
	\begin{tabularx}{\linewidth}{ >{\centering\arraybackslash}X 
			| >{\centering\arraybackslash}X 
			| >{\centering\arraybackslash}X |  >{\centering\arraybackslash}X }
			Features & \textsc{Study 1} & \textsc{Study 2}& \textsc{Study 3}\\
			\hline
			Goal & Study the reproducibility of wave speed measurements and perform an arterial occlusion. & Wave speed and anatomy measurements, comparison with pQCT. & Perform a tilt experiment while imaging blood flow. \\
			Number of volunteers& 12 & 16 & 11 \\
			Ages (y. o)& 
			[20-40]& Two age groups: 20-40 and 50-70&  [24-61]\\
			Number of visits& 2 & 1& 1 \\	
			Figures & 3 \& 4 & 2 \& 3 & 4 \& 5\\
			Clinical trial number &NCT06206031	&NCT06206031&NCT04396288\\		
			\rev{Promoter} & \rev{CHU Angers} & \rev{CHU Angers} & \rev{INSERM C19-30 ERVIO} \\
			\rev{Approval number} & \rev{23.03951.000199-MS01} & \rev{23.03951.000199-MS01} & \rev{20.03.02.58205} \\
			Written informed consent & \checkmark & \checkmark&\checkmark \\	
		\end{tabularx}
	\end{table}
\paragraph{Ultrasound acquisitions}
Acquisitions were performed using a Vantage 256 mid-frequency scanner (Verasonics Inc., Kirkland, WA, USA) connected to clinical cardiac phased-array transducers (see details in Table~\ref{USparameters}). The probe was hand-held and positioned on the anteromedial surface of the mid-diaphysis of the tibia. 2D transverse images were acquired. In a transverse plane, $C^{bone}(\theta) = C^{radial}$ since the image plane is considered normal to the symmetry axis. Acquisition parameters are summarized in Table~\ref{USparameters}.  

\begin{table}
	\caption{Ultrasound and reconstruction parameters}
	\label{USparameters}
	\setlength{\tabcolsep}{3pt}
	\renewcommand{\arraystretch}{1.2} 
	\begin{tabularx}{\linewidth}{| >{\centering\arraybackslash}c 
			| >{\centering\arraybackslash}X 
			| >{\centering\arraybackslash}X |}
			\hline
			Parameters& 
			\textsc{Studies 1 \& 2} 
			& \textsc{Study 3} 
			\\ 
			\hline
			\hline
			\multicolumn{3}{|c|}{\textbf{Ultrasound transducer characteristics} }\\
			\hline
			\hline
			Probe& GE M5Sc-D &   Philips P4-1\\
			Frequency (MHz)& 2.8 & 2.5\\
			Pitch (mm)& 0.23 & 0.295 \\
			Nb elements& 80 & 96  \\
			Lens wave speed (m/s)& 1036  & 970 \\		
			Lens thickness (mm)&0.921 & 1.3 \\
			\hline
			\hline	
			\multicolumn{3}{|c|}{\textbf{Blood flow acquisition parameters} }\\	
			\hline
			\hline	
			Steering angles in skin& $\pm$[12.2;14.9;17.6]\textdegree  &  $\pm$[12.9;15.4;18]\textdegree  \\	
			Framerate (Hz)& 400 & 440  \\	
			Acquisition duration &\qty{4.45}{\second} (1780~frames)& \qty{4.74}{\second} (2088~frames) \\
			\hline
			\hline		
			\multicolumn{3}{|c|}{\textbf{Blood flow reconstruction parameters} }\\
			\hline
			\hline	
			Pixel size (mm)  & 0.158 & 0.158  \\
			F-number  & 1.3 & 1.3  \\	
			\textbf{Transmit compounding} &  & \\
			average angles in skin &  [-14.9\textdegree  ; 14.9\textdegree]  & [-15.4\textdegree ;  15.4\textdegree] \\
						
			\textbf{Receive angles at pixel}& & \\
			 in skin and bone cortex & [-35\textdegree ; 35\textdegree] & [-35\textdegree ; 35\textdegree] \\	
			 in marrow & [-12\textdegree ; 12\textdegree] & [-12\textdegree ; 12\textdegree] \\				

			\hline
		\end{tabularx}
	\end{table}
\paragraph{Lens characterization}
To perform ray tracing accurately in a layered medium, the acoustic properties of the probe itself must be characterized, in particular the rubber lens covering the piezoelectric elements. Hence, the lens wave speed and thickness were estimated using the method proposed in \cite{waasdorp_assessing_2024}. This lens characterization was carried out for both probes used in the study, and the resulting parameters are reported in Table~\ref{USparameters}.

\paragraph{US/pQCT image registration}
Transverse slices of the mid-tibia were acquired using a XCT-3000 pQCT (Stratec Medizintechnik, Pforzheim, Germany) scanner (Fig. \ref{Fig-Anatomy}(a)). This modality provides quantitative images of bone structure and bone mineral density. Fig. \ref{Fig-Anatomy}(b) shows an exemplary pQCT slice (slice thickness was \qty{2}{\milli\metre}), with both bones of the lower leg (tibia and fibula) clearly visible. At each \qty[parse-numbers=false]{0.5 \times  0.5}{\milli\metre\squared} voxel, the bone mineral density in \unit{\gram\per\centi\metre\cubed} is measured. From the pQCT image, the outer cortical interface (periosteum) of the tibia was segmented using a threshold of \qty{710}{\milli\gram\per\centi\meter\cubed} \cite{ward_recommendations_2005}.

The ultrasound images in the same transverse plane (Fig. \ref{Fig-Anatomy}(c)) were reconstructed using the bone-corrected method described in Section~\ref{subsec:reconstruction}). The medium was modeled as three distinct layers: skin, cortical bone and marrow. For improved accuracy, the ultrasound wave speed in each of the three layers was determined beforehand using methods described in section~\ref{sec:WaveVelocityMeasurement} below. The periosteum interface extracted from the ultrasound image was then co-registered with its pQCT counterpart using Arun's least squares rigid transformation algorithm \cite{arun_least-squares_1987}. \rev{During the acquisition, a line was drawn on the skin of the volunteer, in order to make sure that the US and pQCT image planes were precisely aligned. The US slice thickness being of \qty{6}{\milli\metre} and the pQCT slice thickness being of \qty{2}{\milli\metre}, with a millimetric precision we can assume that the two planes are well-aligned.}

\paragraph{Cortical thickness measurement}
After image registration, cortical thickness was estimated independently from both the ultrasound and pQCT images. Within the overlapping region between the two modalities, a second-order polynomial was fitted to the segmented periosteum and endosteum contours (i.e., the outer and inner cortical interfaces closest to the ultrasound probe). The median line between these two curves was then computed. Along the direction normal to this line, the local thickness was measured across the image width by computing the distance between the periosteum and endosteum \cite{renaud_measuring_2020 }. The final cortical thickness was obtained by averaging these local thickness values. Whenever the endosteum was not well-defined on pQCT images, the volunteer was excluded for cortical thickness measurement.

\subsubsection{Results: US/pQCT anatomy agreement}
\begin{figure}[htbp]
	\centering
	\includegraphics[width=\linewidth]{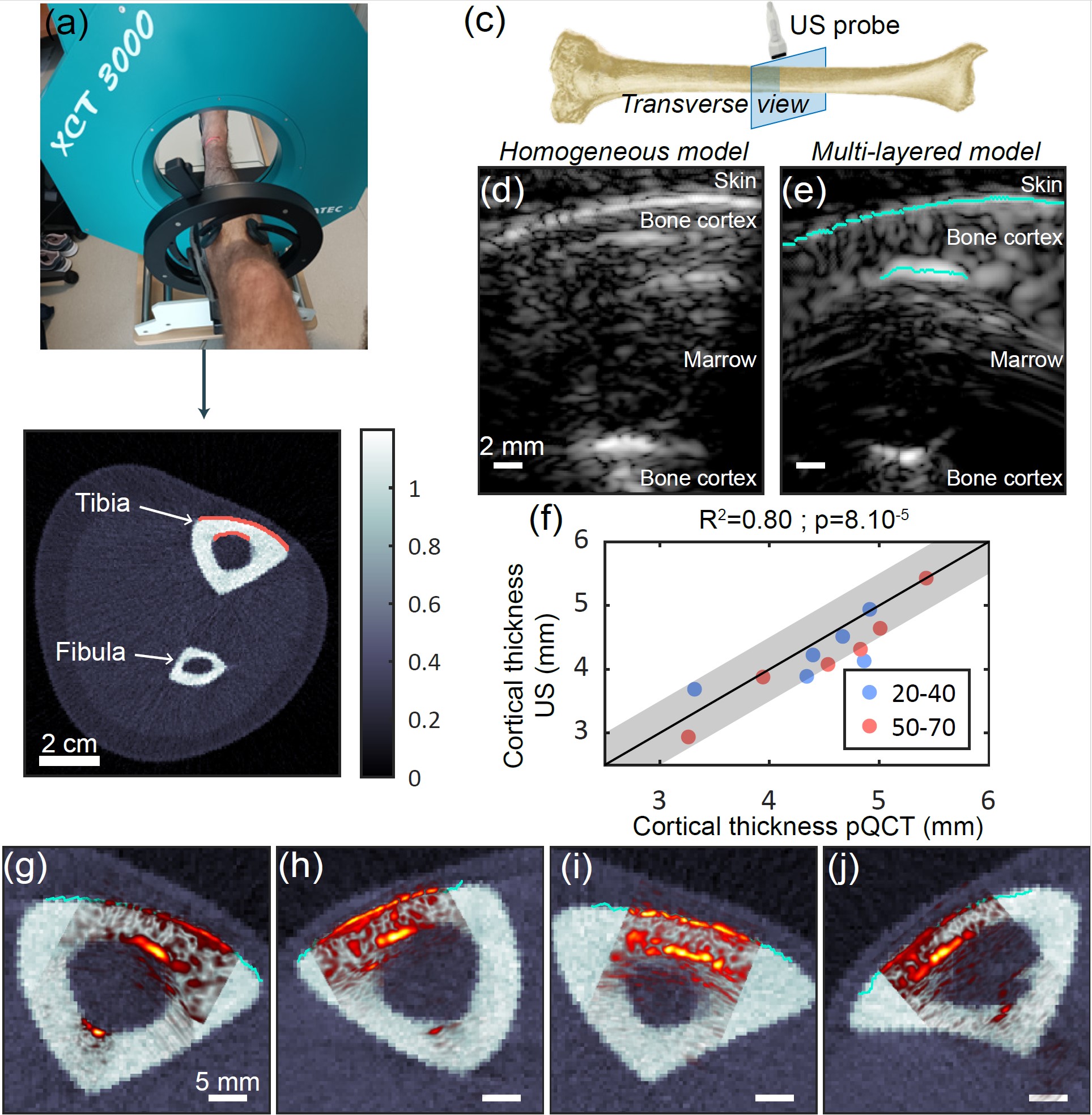}
	\caption{(a) View of the pQCT scanner used to acquire transverse reference images of the tibia.
		(b) Example of a pQCT slice showing the tibial and fibular cortices.	
		(c) Schematic representation of the transverse imaging plane.
		(d) Ultrasound B-mode image reconstructed under the assumption of a homogeneous medium and (e) using the proposed multi-layered model including lens, skin, anisotropic cortical bone, and marrow.
		(f) Cortical thickness measured on the ultrasound images as a function of cortical thickness measured on pQCT slices. 
		(g–j) Superposition of pQCT and ultrasound images for four different volunteers. The outer cortical interface (periosteum) segmented from the ultrasound image is overlaid in cyan on the grayscale ultrasound image.}
	\label{Fig-Anatomy}
\end{figure}

Fig.~\ref{Fig-Anatomy}(d) shows the ultrasound image reconstructed under the assumption of a homogeneous medium. In contrast, Fig.~\ref{Fig-Anatomy}(e) shows the result obtained using the proposed multi-layer ray tracing reconstruction.
Aberration correction clearly improves image quality, enhancing the visibility of both the periosteum and endosteum. For all the 8 young volunteers and 6 out of 8 older volunteers where the endosteum was visible, cortical thickness was measured on US and pQCT images. The result is shown in Fig.~\ref{Fig-Anatomy}(f), showing good correlation ($R^2=0.80$) between cortical thicknesses measured using both imaging modalities. Four volunteers in total had to be discarded because their endosteum was not well defined (two from the younger group and two from the older group). \rev{Indeed, for some volunteers, some trabecularisation of the cortical bone at the endosteal side appeared. In this case, the endosteum is not a clear interface anymore but the transition becomes more progressive, making the cortical thickness poorly defined.} The gray zone corresponds to a \qty{0.5}{\milli\metre} error. Fig.~\ref{Fig-Anatomy}(g-j) present co-registered images from four different volunteers, showing the ultrasound reconstruction (in red) overlaid with the pQCT slice (in black and white). The cyan line correspond to the periosteum segmentation obtained from the ultrasound image and used for image registration. Across all examples, the periosteum and endosteum align well between modalities, confirming the reconstruction accuracy. When the second endosteum is visible (Fig.~\ref{Fig-Anatomy}(g-h-j)), it consistently appears at the correct depth. Minor rotational mismatches are observed in panels (h) and (j), likely due to small errors during the image registration step. \rev{Indeed, when the tibia shape is locally simple, registration is challenging because there is no outstanding geometrical feature to help the registration algorithm.} These observations support the validity of the proposed framework for producing geometrically accurate, bone-corrected ultrasound images. 

\section{Measurement of ultrasound wave speed in bone tissue}
\label{sec:WaveVelocityMeasurement}
\color{black}
In addition to anatomical imaging, the proposed ultrasound framework enables the extraction of quantitative biomarkers related to bone tissue quality. By measuring ultrasound wave speeds in both radial and axial directions, it is possible to assess cortical bone anisotropy, which is closely linked to its elastic and structural properties \cite{granke_change_2011,renaud_measuring_2020}.

\begin{figure*}[ht]
	\centering
	\includegraphics[width=\linewidth]{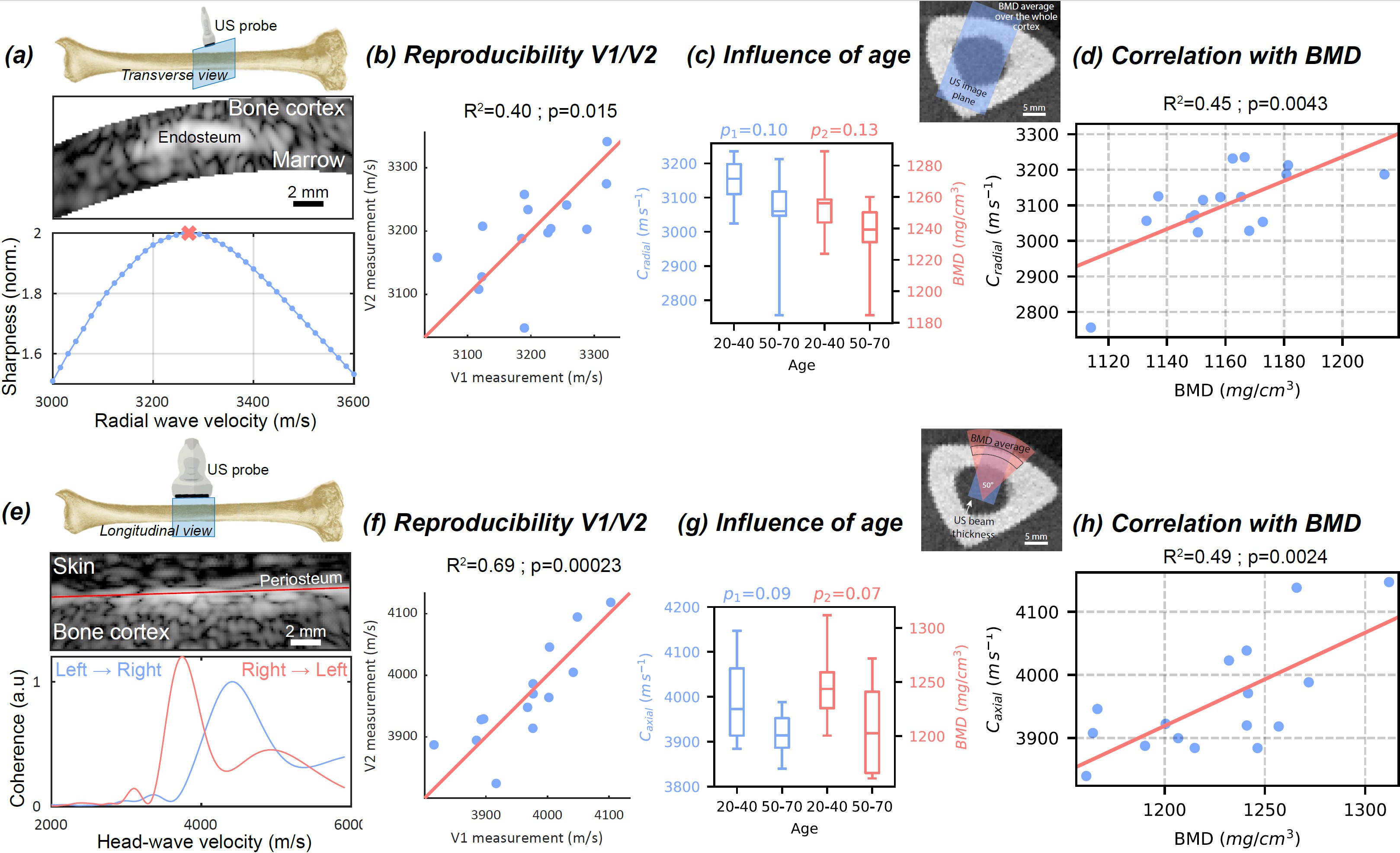}
	\caption{In vivo measurement of radial and axial wave speed in the tibial cortex.
		(a–e) Estimation of radial wave speed using auto-focusing (a) and axial wave speed using head-wave tracking (e).		
		(b–f) Reproducibility between visits.
		(c–g) Wave speed measurements across two age groups.
		(d–h) Correlation between ultrasound wave speed and bone mineral density (BMD) from pQCT.}
	\label{Fig-VelocityMeas}
\end{figure*}

\subsection{Methods for ultrasound wave speed estimation}
\subsubsection{Auto-focusing for radial wave speed measurement}
\color{black}
Radial wave speed was estimated from a transverse view of the tibial diaphysis, using synthetic aperture ultrasound acquisitions. An auto-focusing approach, as previously described in \cite{renaud_vivo_2018, renaud_measuring_2020}, was used to determine the wave speed that maximized a focus metric within the cortical layer. This method relies on iteratively adjusting the assumed wave speed in the cortex and evaluating the sum of two sharpness metrics (Brenner and Normalized Variance). The radial wave speed corresponds to the speed value that yields the highest sharpness value, as shown in Fig.~\ref{Fig-VelocityMeas}(a). For each volunteer, five successive acquisitions were performed at the same site to improve measurement robustness. Due to manual probe positioning, the transverse alignment of the imaging plane may vary slightly. As radial wave speed is the lowest among all propagation directions, the minimum value across the five acquisitions was retained for analysis. 

\subsubsection{Axial wave speed measurement with the head-wave}
Axial wave speed was estimated from a longitudinal view of the tibial cortex, through measuring the speed of a wave guided along the skin–cortex interface (periosteum). This wave, often referred to as a “head wave” can be tracked across successive receive channels. If the wavelength is smaller than the cortical thickness, it propagates at the axial wave speed in the bone cortex \cite{foldes_quantitative_1995}. Following the method introduced in \cite{renaud_measuring_2020}, the wave fronts propagating both from left to right and right to left were tracked in the spatiotemporal ultrasound data. In longitudinal view, the periosteum is locally flat, resulting in a plane head wave front. Its propagation speed in the imaging plane was then computed by linear regression of arrival times across channels (\rev{see Fig.~\ref{Fig-VelocityMeas}(e)}). \rev{Along the interface, the head-wave travels at the same speed in both directions, but since the periosteum might not be parallel to the probe, the "apparent" velocity measured along the transducer elements depends on the propagation direction. Measuring the wave speed in both directions allows to take the angle into account and then obtain an angle-independent measurement \rerev{\cite{bossy_bidirectional_2004}}.} As expected from anisotropic cortical bone properties, the axial wave speed ($C_{axial}$) is typically higher than the radial wave speed ($C_{radial}$) and reflects the longitudinal stiffness of the tissue. As with the radial wave speed measurement, five repeated acquisitions were performed with probe repositioning. Since axial wave speed corresponds to the fastest propagation direction, the maximum value across acquisitions was taken as the best estimate \cite{foldes_quantitative_1995}.

\subsection{Studies design}
This section reports results from two complementary studies focusing on anisotropic wave speed measurements. Study 1 was used to assess measurement reproducibility, as each volunteer underwent two acquisitions spaced 5 to 10 days apart. \rev{The first and second acquisitions performed on each volunteer are called V1 and V2, respectively (for Visit 1 and Visit 2).}
Study 2 aimed to evaluate the influence of age on wave speed and to examine correlations between ultrasound-based speed biomarkers and BMD measured from pQCT. \rev{Studies 1 and 2 were promoted by CHU Angers, the approval number by appropriate ethical committee was 23.03951.000199-MS01 and the clinical trial number was NCT06206031.}

\subsection{Bone mineral density measurement using pQCT}
pQCT provides high-resolution (\qty{0.5}{\milli\metre}) transverse maps of bone mineral density (BMD) across the entire slice. \rev{The imageJ plugin called ``pQCT" \cite{rantalainen_open_2011,rantalainen_differential_2013} was used to calculate the BMD average over specific regions of the cortex.} For comparison with radial wave speed, the BMD was averaged over the full cross-section, as the transverse ultrasound acquisitions cover a wide field-of-view and include the entire cortical thickness. In contrast, axial wave speed is sensitive to the superficial region of the cortex directly beneath the probe, over a thickness of half a wavelength \cite{bossy_three-dimensional_2004}. \rev{In addition, since a longitudinal ultrasound image is used in this case, the ultrasound and pQCT imaging planes are orthogonal. Hence, the overlapping region between the two modalities correspond to the ultrasound plane thickness, which is of \qty{6}{\milli\metre} in our case.} \color{black}Therefore, to match the axial measurement region, BMD was averaged within a 50° polar sector centered on the probe position and limited to the outer third of the cortex. \rev{In Fig.~\ref{Fig-VelocityMeas}(d) and (h), the ultrasound imaging planes are superimposed over a pQCT slice, to show how they relate to one another.} 

\subsection{Results}
Fig.~\ref{Fig-VelocityMeas} summarizes the results obtained for wave speed measurements. The reproducibility results from Study 1 are shown in Fig.~\ref{Fig-VelocityMeas}(b) and (f). Axial wave speed measurements showed higher reproducibility ($R^2=0.69$) than radial measurements ($R^2=0.40$). The approximate placement of the probe at the mid-diaphysis may have impacted reproducibility. Fig.~\ref{Fig-VelocityMeas}(c) and (g) present wave speed measurements from the two age groups in Study 2. As expected, median values are lower in the older group. However, the differences did not reach statistical significance. This lack of significance is likely attributable to the limited sample size (n = 8 per group). Finally, Fig.~\ref{Fig-VelocityMeas}(d) and (h) show the relationship between ultrasound wave speed and pQCT-derived BMD. Significant correlations were observed in both directions, with p-values $<$0.005, $R^2=0.45$ for radial wave speed and $R^2=0.49$ for axial wave speed. These findings are in line with previous \emph{ex vivo} studies on the radius, which reported correlations of $R^2 = 0.49$ between radial wave speed and porosity \cite{eneh_effect_2016}, and $R^2 = 0.57$ between axial wave speed and BMD \cite{bossy_vitro_2004}. \emph{In vivo} studies have reported lower correlations, such as $R^2 = 0.40$ between BMD at the radius and axial tibial wave speed on 307 women \cite{foldes_quantitative_1995}, and $R^2 = 0.29$ between tibial cortical density and axial wave speed in 51 women \cite{sievanen_ultrasound_2001}. Despite a smaller sample size, the correlations observed here are consistently higher than previously reported \emph{in vivo} and comparable to \emph{ex vivo} results. This improvement is likely due to the imaging-based nature of our method unlike earlier blind transmission devices.

\section{Quantitative blood flow imaging inside bone}
In addition to anatomical and tissue structural biomarkers, the proposed ultrasound framework also enables the detection and mapping of intraosseous blood flow, providing insight into the vascular function of cortical bone. We will focus only on Power Doppler imaging, because it was shown to be proportional to the flowing blood volume \cite{rubin_fractional_1995}.

\subsection{Improved extraction of blood flow signal}
\rev{The wave speed model used for blood flow imaging was specific to each volunteer and determined using the methods described beforehand (see section~\ref{sec:WaveVelocityMeasurement}). Then, the  interfaces segmentation was obtained from an accurate anatomical image acquired with the SA sequence, and later used to reconstruct corrected images from the diverging or plane waves sequence.} Finally, to obtain Power Doppler images, a conventional method would be to apply a clutter filter to remove the stationary tissue signal and enhance the moving blood signal. However, in the case of intra-osseous blood flow, this simple approach is not sufficient due to a very small blood volume and very slow blood flow. Therefore, a ray selection strategy was introduced prior to clutter filtering to favor spherically re-radiated signals from red blood cells over specular reflections from the periosteum and endosteum. 

\subsubsection{Methods and ray selection strategy}
The strong acoustic impedance mismatch at the skin/bone (periosteum) and bone/marrow (endosteum) interfaces (a factor of 4) results in highly reflective surfaces, appearing very bright on ultrasound images. Even though these interfaces are nearly stationary, their strong reflections are not well suppressed by the clutter filter. Applying our phase-aberration reconstruction followed by high-pass filtering results in Power Doppler images dominated by bright interface signals (Fig.~\ref{Fig-PowerDopplerImaging}(h)). To address this, a new beamforming approach was developed. It relies on the principle that when an ultrasound wavefront encounters a point target (red blood cells in our case), it re-radiates energy in all directions (omnidirectional diffuse scattering), whereas interfaces reflect energy at a mirrored angle (directional specular reflections). Hence, \rev{data was beamformed with a strategy controlling the F-number, and the reception and transmission angles. Indeed, the F-number was maintained constant (equal to 1.3) throughout the image. Moreover, in reception, the average angle of the rays at all pixels was kept constant in each of the layers: \qty[parse-numbers=false]{\pm 35}{\degree} in the skin and bone cortex layers, \qty[parse-numbers=false]{\pm 12}{\degree} in the marrow layer (see table~\ref{USparameters}). In addition, transmit sub-compounding of the six transmitted angles was performed, with two sub-groups of angles being compounded (see table~\ref{USparameters}). These two sub-compounding groups correspond to two transmission angles. Finally, ray selection relying on having reception and transmission angles of the same sign, in the skin and bone cortex layers images beamformed with \qty{-35}{\degree} reception angle and first transmit angle (-14.9\textdegree\ or -15.4\textdegree) was incoherently compounded with \qty{+35}{\degree} reception angle and last transmit angle (+14.9\textdegree or +15.4\textdegree). In order to have an image that is wide enough, in the marrow layer the reception angles had to be smaller. Therefore, images in this layer were obtained by incoherently compounding the images obtained with: \qty{-12}{\degree} reception angle and first transmit angle (-14.9\textdegree\ or -15.4\textdegree) incoherently compounded with \qty{+12}{\degree} reception angle and second transmit angle (+14.9\textdegree\ or +15.4\textdegree). Ray tracing patterns shown in Fig.~\ref{Fig-PowerDopplerImaging}(a-c) illustrate this principle of having transmit and reception angles of same signs.} 


\subsubsection{Evaluation on phantom experiments}
\rev{To evaluate the method, we conducted phantom experiments using the same acquisition parameters than studies 1 and 3. In a water tank, a Sawbones (Pacific Research Laboratories, Vashon, WA) phantom (\qty{4.4}{\milli\metre}-thick plate mimicking cortical bone) was placed above a nylon wire, and imaged using the GEm5Scd probe. The ray selection reconstruction scheme was applied, selecting first only rays coming from the right (both in transmission and reception), as shown in Fig.~\ref{Fig-PowerDopplerImaging}(a), resulting in a tilted PSF (see Fig.~\ref{Fig-PowerDopplerImaging}(b)). In a second step, rays coming from the left are selected (see Fig.~\ref{Fig-PowerDopplerImaging}(c)), resulting in a PSF tilted in the other direction, as shown in Fig.~\ref{Fig-PowerDopplerImaging}(d). Finally, these two images are incoherently compounded, resulting in the final image using the ray selection approach, displayed in Fig.~\ref{Fig-PowerDopplerImaging}(e). The multi-layered reconstruction obtained similarly than Fig.~\ref{Fig-Methods}(h) was also computed and is shown in Fig.~\ref{Fig-PowerDopplerImaging}(f). In Fig.~\ref{Fig-PowerDopplerImaging}(f) the plate interfaces are clearly visible, as well as the PSF corresponding to the nylon wire, at a depth of \qty{15}{\milli\metre}. However, in Fig.~\ref{Fig-PowerDopplerImaging}(e), the PSF is still visible at the same depth but the interfaces almost disappeared from the image. The average amplitudes along the interfaces were computed and reported on the images directly. The top and bottom interface signals (specular scattering) have been reduced by 9.1~dB and 22.5~dB compared to the amplitude at the nylon wire. Moreover, the bright horizontal lines at 12 and 14~mm, corresponding to multiple reflections either inside the plate or in between the probe and the plate, have been drastically reduced. For example, the bright line at 14~mm (corresponding to the bottom surface image when a multiple reflection happens in between the probe and the top surface) has been reduced by 15.9~dB. Finally, the lateral resolution is degraded by this ray selection method, the full-width-at-half-maximum being increased: $FWHM=0.79~mm$ for conventional bone-corrected reconstruction and $FWHM=1.79~mm$ for the ray selection image. 
}

\subsubsection{\emph{In vivo} imaging of the intra-osseous blood flow}
\begin{figure}[htbp]
	\centering
	\includegraphics[width=\linewidth]{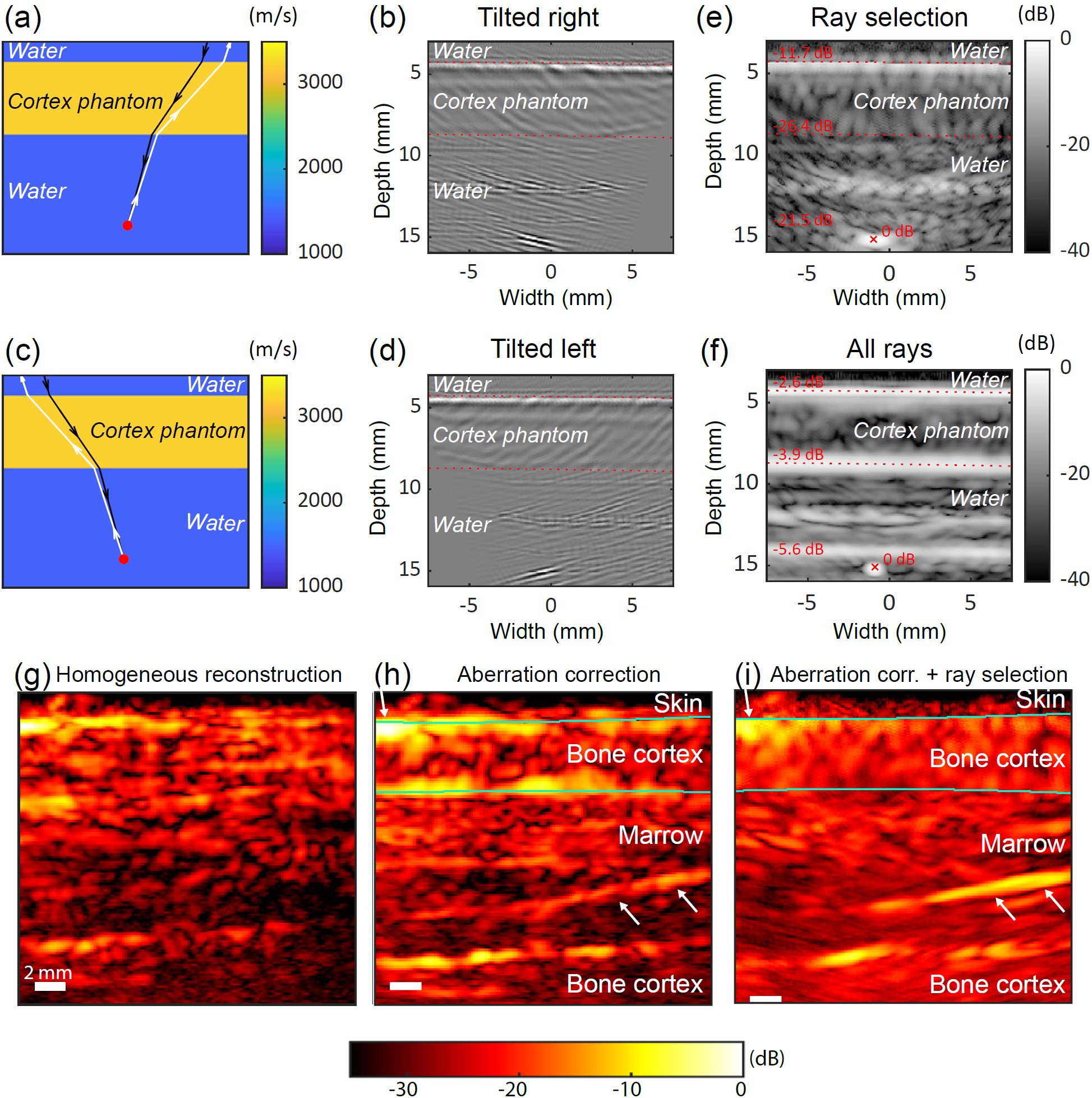}
	\caption{\textbf{Ray selection method and intraosseous blood flow imaging.}
		\rev{
		(a) Ray tracing strategy for the PSF tilted on the right side, and (b) the corresponding image.
		(c) Ray tracing strategy for the PSF tilted on the left side, and (d) the corresponding image.
		(e) Final image obtained using ray selection.
		(f) Image obtained using the aberration correction framework presented above.
		\rerev{(g)} \emph{In vivo} Power Doppler image of the tibial cortex assuming an homogeneous wave speed model, (h) applying the aberration correction and (i) applying the ray selection and aberration correction methods.
		Movie S1 provided as a supplementary material shows the temporal evolution of panel (i) and plots of Power Doppler signals over time in some specific regions..}}
	\label{Fig-PowerDopplerImaging}
\end{figure}

\rev{After beamforming data with controlled angles but before incoherently compounding images,} a fourth-order Butterworth temporal band-pass filter (8-\qty{30}{\hertz}) was applied to further isolate the blood signal. This frequency band corresponds to flow velocities between 4.8 and \qty{18}{\milli\metre\per\second}, consistent with values reported in brain capillaries \cite{santisakultarm_vivo_2012}. \rev{SVD clutter filtering was also tested on our data but proved inefficient in separating the blood signal from the tissue signal, as both exhibit similar speckle patterns. The result obtained by processing data of Fig.~\ref{Fig-PDModulation}(b) (blue and red time points) when applying a SVD clutter filter are shown in supplementary material (Movie S3). All the temporal singular vectors showing pulsatility and a weak DC offset have been selected (4 to 8). The factor between the peak Power Doppler metric during and right after occlusion is reduced from 20 with a band-pass clutter filter to 1.3 with a SVD clutter filter. Moreover, the pulsatility of the Power Doppler signal average over the bone cortex is less regular and more noisy with the SVD clutter filter. We believe that the failure of the SVD filter on our data is due to the similarity of the tissue speckle and the blood speckle in cortical bone. At this ultrasound frequency (2.5-2.8 MHz), intra-cortical pores and intra-cortical blood vessels are not resolved. Tissue speckle is mainly caused by intra-cortical pores. Blood speckle is generated by small vessels inside intra-cortical pores, thus following the same trajectories, and generating very similar speckle.} The obtained Power Doppler images, acquired during systole, are shown in Fig.~\ref{Fig-PowerDopplerImaging}. It compares three reconstruction methods: assuming a homogeneous medium (Fig.~\ref{Fig-PowerDopplerImaging}(g)), correcting for anisotropic bone aberration (Fig.~\ref{Fig-PowerDopplerImaging}(h)), and applying ray selection in addition to aberration correction (Fig.~\ref{Fig-PowerDopplerImaging}(i)). In the homogeneous reconstruction, only the periosteum appears clearly, while deeper structures are not visible due to strong aberrations. With aberration corrections done, interfaces such as the periosteum and endosteum interfaces become prominent, appearing brighter than the surrounding tissue. However, prior histological studies suggest a relatively homogeneous vascularization of the cortical bone \cite{kelly_blood_1961}, which is inconsistent with such interface-dominated signals. In the Power Doppler image, a suspected trans-cortical vessel appears on the left side of the image, overlapping with the periosteum signal. A second vessel is also visible within the medullary canal. After applying ray selection, interface brightness is significantly reduced. The trans-cortical vessel becomes more distinct, no longer confounded with the periosteum. A video of this dataset reconstructed with ray selection is provided as a supplementary material (Movie~S1).
Overall, ray selection leads to a more uniform Doppler signal within the cortical bone and effectively suppresses interface artifacts. This improvement is attributed to the suppression of specular reflections, which typically dominate at large impedance mismatching boundaries and are difficult to eliminate via filtering only. By favoring diffuse scattering from blood cells, this method enhances vascular contrast and reduces interface signals from stationary tissue. \rev{Moreover, even if there might be some multiple reflections in the skin appearing in the bone cortex region, the Doppler signal is higher in the bone cortex (-22~dB) than in the skin (-29~dB). Since multiple reflections are of much lower amplitude than simple reflections, the part of the Power Doppler signal in the bone cortex that can be attributed to multiple reflections from the skin is considered negligible.} Additionally, horizontal artifacts observed in the medullary region—likely due to multiple reflections—are attenuated. Some speckle now becomes visible in the marrow, suggesting that reducing specular reflection also mitigates reverberation artifacts.

\begin{figure*}[htbp]
	\centering
	\includegraphics[width=\linewidth]{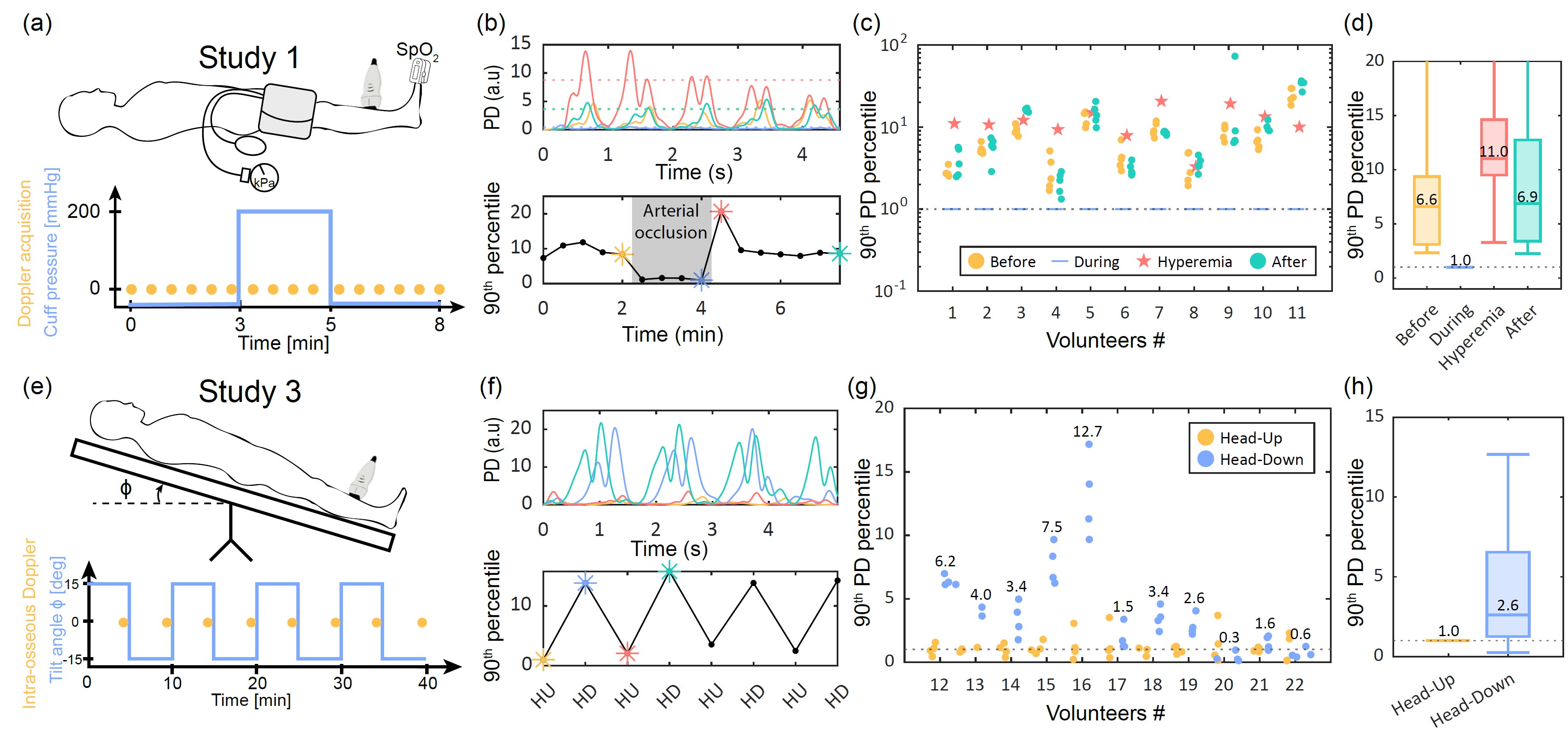}
	\caption{\textbf{Modulation of intraosseous blood flow using Power Doppler imaging.}
		(e) Timeline of the occlusion experiment with Doppler acquisitions.
		(f) Power Doppler (PD) signals during occlusion and hyperemia for one volunteer. Dashed lines show the 90$^{th}$ percentile.
		(g) Summary of the occlusion experiment results across all volunteers.
		(h) Aggregated data showing median ratios between different experimental phases.
		(i) Timeline of the tilt experiment with Doppler acquisitions.
		(j) Power Doppler signals acquired in head-up (HU) and head-down (HD) positions for one volunteer.
		(k) Summary of the tilt experiment results across all volunteers.
		(l) Aggregated data showing the overall effect of tilt on intra-osseous blood flow.}
	\label{Fig-PDModulation}
\end{figure*}

\subsection{Sensitivity to modulation of flowing blood volume}
To further validate intra-osseous blood flow imaging, Studies 1 and 3 included physiological and non-physiological modulation protocols.

\subsubsection{Studies design}

\paragraph{Arterial occlusion}
In Study 1, the volunteer was placed in a supine position. A pressure cuff was positioned on the thigh, and the ultrasound probe was fixed at the mid-diaphysis of the tibia using a dedicated probe holder (USONO, Netherlands). A complete arterial occlusion was achieved by inflating the cuff to \qty{180}{\milli\meter Hg}, effectively blocking the femoral artery and associated vessels. An automated Hokanson device was used to rapidly inflate the cuff and maintain the occlusion for 2~minutes before sudden deflation. Throughout the experiment, \qty{4.45}{\second} Doppler acquisitions were performed every \qty{30}{\second} to monitor intra-osseous blood flow over time. The experimental timeline is illustrated in Fig.~\ref{Fig-PDModulation}(a) and ultrasound acquisition parameters are reported in Table~\ref{USparameters}. \rev{Study 1 was promoted by CHU Angers, the approval number by appropriate ethical committee was 23.03951.000199-MS01 and the clinical trial number was NCT06206031.}

\paragraph{Head-up/head-down tilt}
In Study 3, a tilt-table protocol was used to modulate tibial blood flow physiologically. The volunteer lay on a tilt table alternating between head-up and head-down positions at \qty{\pm 15}{\degree}. The experiment comprised eight 5-minute phases, switching between inclinations. After each 5-minute stabilization period, a \qty{4.74}{\second} Doppler acquisition was performed (see Table~\ref{USparameters}) just before changing position. The ultrasound probe was fixed at the mid-diaphysis of the tibia using a dedicated probe holder (USONO, Netherlands). The procedure is summarized in Fig.~\ref{Fig-PDModulation}(e). \rev{Study 3 was promoted by INSERM through protocol C19-30 ERVIO, the approval number by appropriate ethical committee was 20.03.02.58205 and the clinical trial number was NCT04396288.} 

The sensitivity of intra-tibial blood flow to tilt-induced changes has been previously demonstrated using optical methods \cite{howden_bone_2017,becker_tibia_2018}, with reported average increases in peak blood flow signal by a factor of 3 to 4 in head-down versus head-up positions. This well-documented physiological response provides a reference to assess the sensitivity of our ultrasound-based method to detect tibial blood flow variations.

\subsubsection{Bone-probe motion estimation}
In addition to the processing steps described above, potential motion of the volunteer relative to the probe was monitored throughout the acquisitions. To quantify this motion, a phase-tracking algorithm \cite{loupas_axial_1995} was applied to estimate the velocity of the periosteum. The algorithm computes the phase shift between two consecutive frames at each pixel. Knowing the spatial period in the image, these phase shifts are converted into local displacement (in meters) and vertical velocity. The resulting displacement field is then averaged over the periosteum region to obtain a single time-resolved estimate of probe-bone axial motion. Whenever the periosteal velocity exceeded a predefined threshold (\qty{60}{\micro\metre\per\second} in study 1 and \qty{150}{\micro\metre\per\second} in study 3), a \qty{2}{\second} time window centered on the motion peak was discarded from the Doppler signals. This motion-rejection procedure was systematically applied to all acquired datasets.

\subsubsection{Results for the occlusion experiment}
Fig.~\ref{Fig-PDModulation}(b) shows four examples of Power Doppler signals averaged over the cortex for one volunteer. The signals correspond to acquisitions just before the occlusion, immediately before cuff release, right after release, and during the recovery phase. \rev{Two peaks are observed at each heartbeat in all cases, this is what is called a biphasic flow. In this case, recording a biphasic flow at the vascular level is considered normal and is due to the vessel’s overall vascular properties, depending on systemic vascular compliance and total peripheral resistance. Indeed, even if intraosseous vascular flow has been very little studied so far, using Laser Doppler technology the vascular flow at the tibial diaphysis was already found to be biphasic \cite{binzoni_pulsatile_2013}.} \color{black} Moreover, a marked amplitude drop is observed during occlusion (blue trace), while the post-occlusion signal (red trace) is approximately three times higher than baseline levels (yellow and cyan traces). \rev{Movie S2 in supplementary material shows Power Doppler videos just before and right after the occlusion release. They exhibit a quasi-uniform bone cortex vascularization, while there is a factor of 20 between the peak Power Doppler signals.} This post-occlusion increase is consistent with reactive hyperemia, previously described in skin and muscle tissues \cite{tierney_adaptive_2017}, where blood flow temporarily exceeds baseline to compensate for the occlusion period. To quantify the changes over time, the 90th percentile of each 4.45s Doppler trace was calculated and plotted across the experiment timeline (Fig.\ref{Fig-PDModulation}(b), lower panel). \rev{This metric was found to be representative of the average peak value of the 4 heartbeats recorded within one single acquisition. In comparison, the maximum would be sensitive to the most extreme value, which could be a motion artifact or a heartbeat with a higher peak value compared to the others.} The signal remained stable before occlusion, dropped tenfold upon cuff inflation, and recovered rapidly after release, with a prominent hyperemia peak.

Fig.~\ref{Fig-PDModulation}(c) summarizes the results from the 11 volunteers whose data were analyzable. Data from two volunteers were excluded due to differing acquisition parameters, and one due to excessive motion artifacts. In yellow are plotted the pre-occlusion measurements, in blue the last acquisition during occlusion, in red the first measurement post-release, and in green the measurements obtained from 1 minute after release. All data were normalized by the last measurement during occlusion. In all cases, the ratio between pre-/post-occlusion signals and those during occlusion was at least two and reached up to 30. A hyperemia peak was detected in approximately half the volunteers (subjects \#1, 2, 4, 6, 7, 9), with values exceeding pre- and post-occlusion baselines. Fig.~\ref{Fig-PDModulation}(d) presents the aggregated data, taking the median across acquisitions for each phase. The median signal ratio between pre-occlusion and occlusion was 6.6, comparable to the 7.1 ratio between post-release and occlusion. Moreover, the median ratio between the immediate post-release signal and the last measurement during occlusion reached 11.1, further confirming the hyperemic response in cortical bone.

\subsubsection{Results for the tilt table experiment}

Data from the tilt experiment were analyzed using the same approach. Fig.~\ref{Fig-PDModulation}(f) presents four Doppler signals: yellow and red traces correspond to head-up positions, while blue and cyan were recorded during head-down tilt.
A clear difference is observed, with the Doppler signal being approximately ten times higher in head-down position. The 90th percentile of all Doppler traces during the experiment is shown below, illustrating consistent alternation between low (head-up) and high (head-down) flow states.
Fig.~\ref{Fig-PDModulation}(g) shows the results for all 11 volunteers, with yellow points corresponding to head-up and blue to head-down positions. All values were normalized by the median of head-up measurements. In 9 of 11 volunteers, intra-osseous blood flow was higher in the head-down position, with the median ratio per volunteer displayed on the graph. Finally, Fig.\ref{Fig-PDModulation}(h) shows the overall effect across the cohort, with a median flow increase of 2.7 between head-up and head-down positions. This result is consistent with previous observations using non-spatially specific near-infrared methods, which reported an average increase by a factor of 3 to 4 \cite{howden_bone_2017,becker_tibia_2018}.

\section{Discussion}

\begin{figure}[h!tbp]
	\centering
	\includegraphics[width=\linewidth]{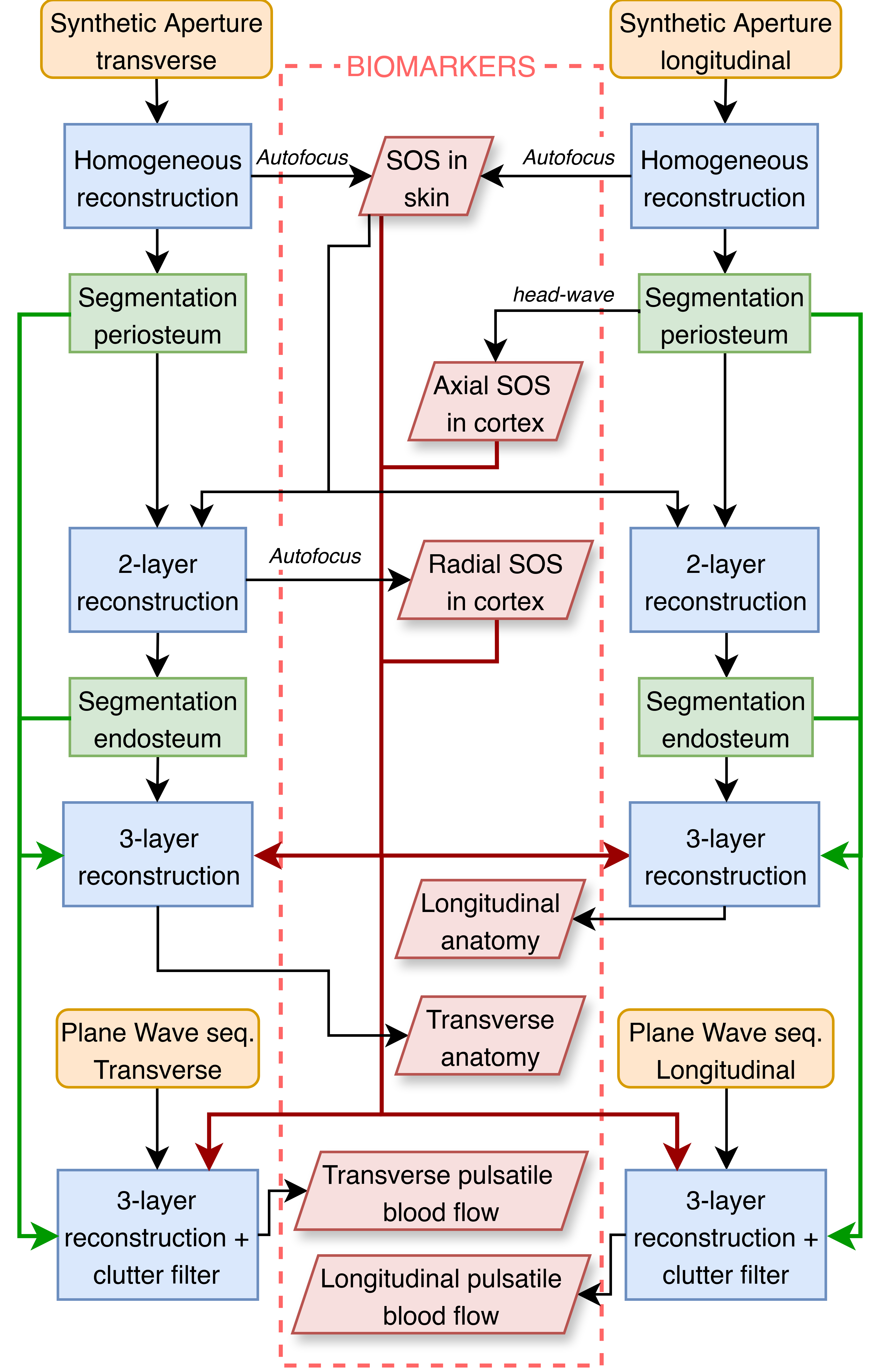}
	\caption{
		\rerev{\textbf{Overview of the bone ultrasound framework developed, allowing to extract different biomarkers.}
The beige boxes correspond to the radiofrequency input data that is obtained from the ultrasound sequences. The blue boxes correspond to reconstruction steps, performed using the aberration correction approach. The green boxes correspond to interfaces segmentations extracted from the images. Finally, the red boxes correspond to clinical biomarkers that can be used to assess  tissue quality through the ultrasound wave speed (SOS=Speed-of-Sound), bone anatomy or intra-osseous blood flow.
}}
	\label{Fig-Flowchart}
\end{figure}

This study presents an ultrasound framework that integrates anatomical imaging, tissue quality assessment, and intra-osseous blood flow mapping for assessing cortical bone. \rerev{The flowchart presented in fig.~\ref{Fig-Flowchart} summarizes the extraction of the different biomarkers and highlights their interdependence. Anatomical reconstructions and blood flow imaging require speed-of-sound estimates, while wave speed measurements, in turn, rely on interface segmentations derived from these reconstructed images.} Anatomical images \rev{of the near cortex previously published \cite{renaud_vivo_2018} were extended to the marrow layer and even the far bone cortex on the other side. These images were validated with pQCT references. Wave speed measurement methods \cite{renaud_vivo_2018} have been shown to significantly correlate with bone mineral density. Finally, intra-osseous Power Doppler maps were obtained for the first time, thanks to the bone-corrected image reconstruction framework and the ray selection method allowing to filter the interface signal to further enhance vessels visualization. Finally, the sensitivity of intra-osseous blood flow to two modulation experiments was demonstrated.}

\paragraph{Applicability to other bone sites}
The tibia was selected due to its superficial position providing easy access for US, and because diaphysis of tibia has nearly cylindrical symmetry. Additionally, the anteromedial surface of the tibial diaphysis is relatively flat, making it particularly suitable for a linear-array transducer. It also helps holding the probe still. Hence, it was ideal for this methodological validation study. For deeper and non-cylindrical bones, a matrix array would be better suited.

Anatomical coverage of our imaging technique remains limited. Only a portion of the long bone cross-section can be visualized in a single ultrasound image. This limitation is exacerbated in our context by strong refraction and specular reflections at bone interfaces. Using transducers with a larger aperture compared to the cardiac probes used in this work could expand the field-of-view and potentially enable full cross-sectional imaging from a single acquisition.

\rev{Finally, at the diaphysis of long bones, the assumption of a medium made of multiple homogeneous layers is reasonable since the cortex is quite homogeneous in a region of a few cm$^2$ \cite{pazzaglia_morphometric_2013}. However, in older volunteers, especially at metaphyseal sites, increased cortical porosity makes the cortico-trabecular transition zone more difficult to visualize. More specifically, there is a porosity gradient (increasing porosity from periosteum to endosteum) which results in a wave speed gradient. In this case, we would need a more continuous wave speed map, and for example an Eikonal solver to model wave propagation.}

\paragraph{Ultrasound wave speed as a bone quality biomarker}
Measured wave speeds showed significant correlations with BMD. In previous studies, axial wave speed was commonly measured using devices such as the Sunlight Omnisense™ system \cite{njeh_assessment_2001,muller_comparison_2005}, which estimates the axial wave speed at the tibia and the radius without anatomical imaging. While such systems have demonstrated clinical utility and reproducibility, the absence of imaging guidance can limit alignment accuracy and may partly explain the lower correlation values reported in those studies ($R^2 = 0.29$ in a cohort of 51 women \cite{sievanen_ultrasound_2001}). In contrast, our approach combines anatomical feedback with wave speed estimation, allowing more consistent probe positioning and interpretation. Other methods were also proposed, based on Lamb guided waves speed measurements. Low-frequency (\qty{200}{\kilo\hertz}) time-of-flight measurements \cite{biver_associations_2019} were shown to correlate with BMD and cortical thickness. A more sophisticated approach based on solving the inverse problem for the multi-modal analysis of guided ultrasound waves allowed for measuring the cortical thickness and a porosity index \cite{minonzio_ultrasound-based_2019}. The present study reports, for the first time, an \emph{in vivo} validation of radial wave speed measurements compared to BMD. This parameter provides complementary insights to axial wave speed, as it is sensitive to a different elastic modulus of the cortical bone, and could therefore offer added diagnostic value in future studies.

\paragraph{Sensitivity to motion and alignment}
Motion artifacts remain a major limitation, especially given the slow velocities of intraosseous blood flow. Even subtle voluntary or involuntary movements can severely degrade signal quality, requiring strict data rejection criteria in this study. In addition, probe alignment is also critical, as accurate imaging depends on acquiring signals within the imaging plane. With linear array transducers, the backscattered echoes—particularly from the periosteum and endosteum—must lie within the 2D acquisition plane. However, due to individual anatomical variability and the non-cylindrical geometry of the tibia, this condition is not always met. Transitioning to matrix-array probes with 3D volumetric acquisition would mitigate alignment sensitivity by capturing the full echo field, improving robustness and ease of use in clinical settings. In addition, it would allow for 3D motion correction.

\paragraph{Detection of hyperemia in cortical bone}
For the first time, reactive hyperemia following occlusion was observed in cortical bone in half of the volunteers. This inconsistency is attributed to limited temporal sampling (one acquisition every 30 seconds), potentially missing peak hyperemic responses occurring typically between 15 and 45 seconds post-release. Increasing sampling frequency post-occlusion would likely capture hyperemia more reliably.

\section{Conclusion}
The proposed ultrasound approach enables, with one single examination and one single device, access to three bone biomarkers: anatomy, tissue quality and blood flow. This study reports the first \emph{in vivo} evaluation with a comparison to peripheral quantitative x-ray computed tomography and modulations of blood flow. Currently, accessing these three bone biomarkers with one single device is not available. Given the proven or suspected role of bone tissue vascularization in bone disorders like osteonecrosis, delayed fracture healing, osteoporosis or osteoarthritis, we believe this non-ionizing and portable approach has the potential to help diagnosis and patient monitoring.


\section*{Acknowledgments}
The authors warmly acknowledge Henry den Bok for his continuous technical support throughout the project. They are also grateful to CNES (Centre National d'Études Spatiales) and in particular to Guillemette Gauquelin Koch, for encouraging this work. The authors further acknowledge the DRCI of CHU Angers and INSERM for their essential administrative assistance in enabling these studies.

\bibliographystyle{ieeetran}
\bibliography{Biblio}

\end{document}